\documentclass[%
 reprint,
 amsmath,amsfonts,amssymb,
 aps,
]{revtex4-2}

\usepackage{algorithmic}
\usepackage{braket}
\usepackage{caption}
\usepackage{csquotes}
\usepackage{graphicx}
\usepackage{mathtools}
\usepackage{subcaption}
\usepackage{url}

\DeclareQuoteStyle{english}%
    {\textquoteleft}
    [\textquoteleft]
    {\textquoteright}
        [0.05em]    
    {\textquotedblleft}
    [\textquotedblleft]
    {\textquotedblright}

\begin{document}
\setquotestyle{english}
\preprint{AIP/123-QED}

\title{Q2Graph: a modelling tool for measurement-based quantum computing}
\author{Greg Bowen}
 \email{gregory.bowen@student.uts.edu.au}
\author{Simon Devitt}%
 \email{simon.devitt@uts.edu.au}
\affiliation{
 Centre for Quantum Software and Information, University of Technology Sydney, Sydney, NSW 2007, Australia
}%

\date{\today}
\begin{abstract}
The quantum circuit model is the default for encoding an algorithm intended for a NISQ computer or a quantum computing simulator.  A simple graph and through it, a graph state - quantum state physically manifesting an abstract graph structure - is syntactically expressive and tractable.  A graph representation is well-suited for algorithms intended for a quantum computing facility founded on measurement-based quantum computing (MBQC) principles.  Indeed, the process of creating an algorithm-specific graph can be efficiently realised through classical computing hardware.  A graph state is a stabiliser state, which means a graph is a (quantum) intermediate representation at all points of the algorithm-specific graph process.  We submit Q2Graph, a software package for designing and testing of simple graphs as algorithms for quantum computing facilities based on MQBC design principles.  Q2Graph is a suitable modelling tool for NISQ computing facilities: the user is free to reason about structure or characteristics of its graph-as-algorithm without also having to account for (quantum) errors and their impact upon state.
\end{abstract}

\maketitle

\section{\label{sec:level1}Introduction}
Quantum computing is an evolutionary step, to the extent that it exhibits solutions to some well-structured computational problems that remain prohibitive, even in theory, to classical facilities.  It is the difference in fundamentals of the two computing models that accounts for much of the distance between classical and quantum computing capabilities.  Unlike the voltages that determine a bit in classical computing, quantum computing derives from properties of an elementary particle and its mathematical abstraction, the \textit{qubit} (Cf. \cite{FMM12}).  As such, a qubit presents as a two-state system with an orthonormal basis, within a complex $2^n$-dimensional vector space.  The qubit as a vector can be manipulated in accordance with standard linear algebra and as implied by its two-state system, a qubit of the \enquote{computational basis} of $\ket 0$ and $\ket 1$ corresponds to the binary bit-state capabilities (i.e. \{0,1\}) of classical computing.  The reader is referred to \cite{NC10} for a comprehensive overview of both the notations and the linear algebra to appear in this article.

The state vector of a qubit's orthonormal basis marks the point of departure of quantum- from classical computing.  Expressing an arbitrary two-state system, $\ket \psi$ in the computational basis gives, \begin{equation}
    \ket\psi = \alpha\ket{0} + \beta\ket{1},
\end{equation}
by which the amplitudes $\alpha$, $\beta$ are complex numbers and by virtue of orthonormality,\begin{equation}
    |\alpha|^2 + |\beta|^2=1.
\end{equation}
This third, \enquote{superposition} state of a qubit as represented by (1) has no equivalent in classical computing insofar as it describes a linear combination of $\ket 0$ and $\ket 1$.

As with superposition, the property of \textit{entanglement} is a feature of quantum computing not found in the binary state of classical computing.  An entangled pair of particles is an inseparable state even after subsequent transformation of the state or physical relocation of either particle.  Of more direct relevance, entanglement of qubits is a precondition to the property of quantum teleportation in computational processes, which is laid out in Section II, below.

The future of quantum computing appears promising although the reality of those quantum computing facilities presently available is more nuanced.  Quantum computing facilities as at December 2021 are classed as noisy intermediate-scale quantum (NISQ) computers and are characterised both by low counts of physical qubits (\textit{n} $\leq 130$) \cite{AAB19,GPS21,WPJ21} and sensitivity to errors arising from environmental decoherence \cite{Pre18,KTC19}.  Furthermore, real-time access to a NISQ computing facility is limited to services provided by a few cloud-based vendors.  The average user must accept the risk of realising an error(s) when processing an algorithm on a NISQ computing facility while accruing a financial cost.

A number of offerings are available to simulate the workings of an idealised quantum computer on a classical computer, commonly in the form of a software development kit (SDK) or a platform as a service (PaaS) offering\footnote{see https://qosf.org/project\_list/ for an extensive list of quantum computing simulator libraries or online resources}.  Due to their ready availability, these \enquote{simulators} are in the forefront of experimentation in quantum computing: the user of a given simulator can expect to prepare the state of (simulated) qubits and transform those qubits by means of standard operators to then observe an output free of errors.  It is noteworthy that the format of passing instructions to a NISQ computer or its simulators is far closer to the punch-cards associated with classical mainframes of the 1940s through 1960s than to the familiar, human-friendly programming languages dominant since the 1970s.  Indeed, a \textit{circuit} format has become the default for encoding an algorithm intended for a NISQ computer or a quantum computing simulator.  To be sure, some simulator offerings have co-opted such programming languages as Python so that the user can organise and approximate its (quantum) computation yet such language extensions are an aid to the (classical computing) user transitioning to the quantum computing paradigm.  The computing model founded on the (quantum) circuit, as proposed by \cite{Fey86} dominates those simulator resources presently available.

A circuit comprises of one or more qubits as input, which are transformed through the use of single- or multi-qubit operators collectively referred to as quantum logic gates; the output qubits are specified or \enquote{measured} (see section 2, below) only once all transformations are completed.  The (quantum) circuit \underline{model} therefore is the method of computation consistent with the circuit as algorithm: as a method, the quantum circuit model \enquote{builds up} output through the aggregated transformations of input qubits.  It is not the case, however, that the circuit model is the sole option for executing a quantum computation: an alternative model in measurement-based quantum computing (henceforth, MQBC) proceeds through single-qubit operations applied in a certain order and basis to an agglomeration of qubits formed through quantum entanglement, known as a \enquote{cluster state} \cite{RHG06}.  In contrast to the circuit model, MBQC can be thought of as \enquote{emerging} the desired output from a cluster state \cite{DKP07}.  The cluster state requires no additional input qubits to establish a (quantum) state \cite{RB02}.  Revisiting its purpose as an algorithm, the circuit is a complicated fit for the MBQC model: a circuit can be adapted to MBQC but only after deconstructing each multi-qubit gate to its single-qubit equivalent then securing the output through measurement.  In brief, the circuit as syntax is inefficient for encoding an algorithm under the MBQC model.

In light of the inefficiency of adapting the circuit to the MBQC model the question becomes, what is an efficient algorithm representation for the MBQC model?  Under the circuit model, individual qubits are the resource from which a computation is built up by means of single- or multi-qubit transformations of individual qubits.  In contrast, MBQC proceeds through single-qubit operations that transform/consume a cluster state resource.  This cluster state is therefore pivotal to the question of efficiency in encoding an algorithm for the MBQC model.  It turns out that a subgroup of the quantum state, known as a \textit{graph state} has a syntax, in the form of graphs, that can model a cluster state.  A graph, as an expression of graph state, can efficiently implement a prescribed set of transformations of a cluster state.  In effect, the graph as an algorithm is to MBQC what the circuit is to the circuit model.

We submit Q2Graph (https://github.com/QSI-BAQS/Q2Graph), a software package for the design and testing of a graph as an algorithm for MQBC.  Q2Graph is a standalone executable and draws upon (simple) graphs as proxies of cluster states to fulfil its requirements as a design tool.  Q2Graph differs from other well-known graph visualisation/design packages (e.g. \cite{BHJ09,Coe18,GN00,HSS08,Sag22,Tan22}) by its primary purpose namely, the design and modelling of a graph as an algorithm for an MBQC system\footnote{The Wolfram Community has proposed ways to configure Mathematica to reproduce a cluster state and transformations of it (https://community.wolfram.com/groups/-/m/t/2499412, accessed 21-AUG-2022).}.  More specifically, the modelling functionality of Q2Graph includes the ability to reproduce the effect of single-qubit operations upon the structure of a graph.  The remainder of this article will serve as a detailed introduction to version 0.1 of Q2Graph.  Section II is a review of MBQC, graphs and graph states, which, together, form the theoretical underpinnings of the graph as a unit of instructions for MBQC.  With the previous section for context, Section III is a review of the design principles that inform Q2Graph, to stimulate potential use cases for the application.  In particular, the algorithm-specific graph is introduced as a model to bridge between classical- and NISQ-computing facilities.  Section IV is a review of Q2Graph functions with an emphasis upon how a user might use the package to model a quantum algorithm.  The article concludes with Section V, including some comments on next steps for development of Q2Graph version 0.2.

\section{Measurement-based quantum computing, graphs and graph states}
Measurement-based quantum computation emerges a computation from a resource of entangled qubits (\enquote{cluster state}) by transforming single qubits, in a certain order and basis.  In preparing the cluster state (denoted as $\ket\Phi_C$), each of its qubits, except those designated as input, are initialised at state $\ket{+}$.  Following on from initialising the component qubits is the process of creating \textit{quantum entanglement}, which is necessary to creating $\ket\Phi_C$.  As implied by its title, $\ket\Phi_C$ is a composite or \textit{multipartite} system of \textit{n}-interacting qubits.  Ordinarily the state space of a composite physical system (say, $\ket{\Psi}$) is obtained as the tensor product of its component qubits, thus, \begin{equation}
    \ket{\Psi}\equiv\ket{\psi_1}\otimes\ket{\psi_2}\otimes...\otimes\ket{\psi_{\textit{n}}}.
\end{equation}
In contrast, applying a controlled-Z (\enquote{CZ}) gate between qubits $\ket{i}_C$ and $\ket{j}_C$ transforms to the state space $\ket{ij}_C$ not $\ket{i}_C\ket{j}_C$ \cite{AMD20,BBB21} as would be consistent with (3).  This \enquote{entangled} state space of $\ket{ij}_C$ is inseparable, there is no observable point of difference in state between its component qubits.  By definition then, \begin{equation}
    \ket\Phi_C \neq \ket{i}_C\otimes\ket{j}_C\otimes...\otimes\ket{\textit{n}}_C.
\end{equation}

Multipartite states like $\ket\Phi_C$ are efficiently described by the stabiliser formalism (see \cite{Got97,Got09,FMM12,Ter15}).  A stabiliser state is an eigenvector of an operator, \textit{U} with eigenvalue +1 (i.e. $\textit{U}\ket\psi=\ket\psi$).  In every instance of a multipartite state, \textit{U} will include operators of the Pauli group, $\mathcal{P}^N\coloneqq\{\pm{\textit{I}},\pm{\textit{iI}},\pm\textit{X},\pm\textit{iX},\pm\textit{Y},\pm\textit{iY},\pm\textit{Z},\pm\textit{iZ}\}^N$.  More will be made of both stabiliser states and $\mathcal{P}$ below, in the context of graphs and graph states.

So much for defining the state space of $\ket\Phi_C$.  Computation in MBQC now proceeds through \textit{measurement}, which was characterised above as emerging a computation from the $\ket\Phi_C$ resource.  The act of measuring a qubit reduces it to one of two states - by convention, $\ket 0$ or $\ket 1$ - with a probabilistic outcome determined by the amplitudes of (1) and irreversibly destroys superposition in the process.  Despite the probabilistic outcome of measuring qubit $\ket\Phi_C^{\textit{j}}$, MBQC is a deterministic computing model: applying local unitary operators (\enquote{corrections}) after completing all measurements will resolve randomness in the computation \cite{RB01,RBB03,BKM07}.

Measuring $\ket\Phi_C$ in a definite order and basis exploits a function of entanglement known as \textit{quantum teleportation}.  Any transforming or measuring of (entangled) qubit A fully and immediately manifests in (\enquote{teleports to}) the state of (entangled) qubit B.  Consequently, measuring qubit ${\ket\Phi}^i_C$ changes the state of $\ket\Phi_C$ to, \begin{equation}
    \ket{\mu}_{\textit{C} \backslash \textit{N}}\otimes\ket{\hat{\Phi}}_\textit{N},
\end{equation}
which is read as the tensor product of state $\ket{\mu}_{\textit{C} \backslash \textit{N}}$, comprising of all measured qubits and state $\ket{\hat{\Phi}}_\textit{N}$, encompassing those qubits yet to be measured (i.e. $\textit{N}\subseteq\textit{C}$) \cite{RB01}.  In effect, computation is the dynamic byproduct of measuring qubits of $\ket\Phi_C$.

With the quantum properties fundamental to the workings of MBQC now summarised, it remains to address the workings of a \textit{graph} as syntax for modelling a quantum state.  A simple graph is a sufficient abstraction to represent a quantum state consisting of either physical qubits or the composite systems known as logical qubits.  As a working definition, a logical qubit is a construct that is created to mitigate the quantum errors that can arise from environmental decoherence (e.g. \cite{Pre98,FMM12,HFD12,Ter15,Lit19}).  Further specifics of logical qubits and fault-tolerant quantum computing are not considered in this paper.

\begin{figure}[h!]
    \centering
    \includegraphics[width=88mm]{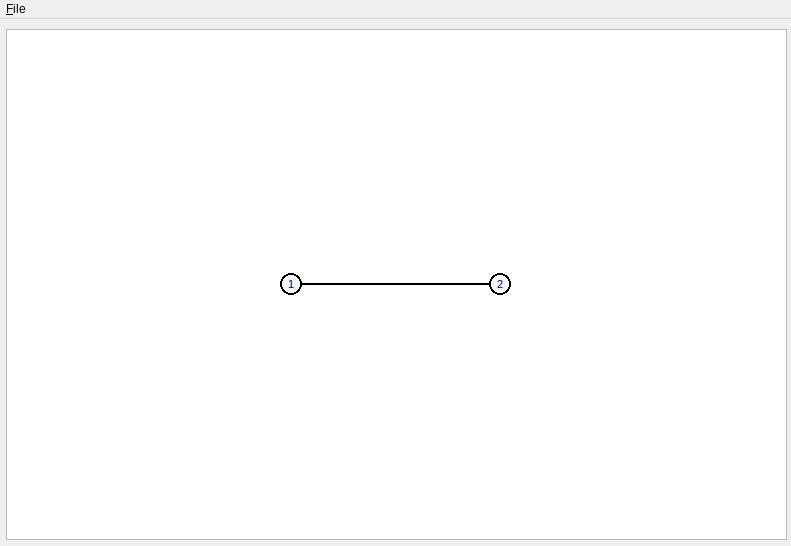}
    \caption{Two vertices connected by an edge.}
    \label{fig:2v_1e}
\end{figure}

A simple graph, \textit{G}, consists of a non-empty, finite set of vertices, \textit{V(G)}, doubling as qubits and a finite set of edges, \textit{E(G)}, which symbolise an interaction between distinct pairs of vertices (qubits).  Note, Q2Graph only deals in simple graphs, which prohibit: 
\begin{enumerate}
    \item a vertex (qubit) joining with itself (a \enquote{loop}); and
    \item joining two vertices (qubits), \textit{a} and \textit{b} with more than one edge.
\end{enumerate} 
In brief, a more expressive graph admits operations which are inconsistent with a stabiliser state.  In Q2Graph, a qubit (vertex) appears as a circle labelled with an identifying integer, while an edge appears as an unbroken line connecting qubit (vertex) \textit{a} and qubit (vertex) \textit{b}, see Figure \ref{fig:2v_1e}.  An edge, \textit{ab}, in Q2Graph signifies an interaction between qubit \textit{a} and qubit \textit{b}.

\begin{figure}[ht]
    \centering
    \begin{subfigure}{0.4\textwidth}
        \includegraphics[width=\textwidth]{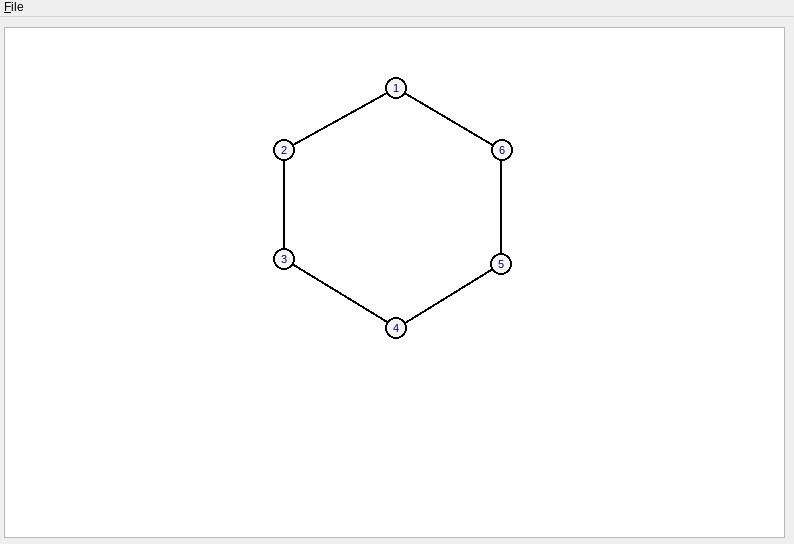}
        \caption{A simple six-vertices graph.}
        \label{fig:6v}
    \end{subfigure}
    \hfill
    \begin{subfigure}{0.4\textwidth}
        \includegraphics[width=\textwidth]{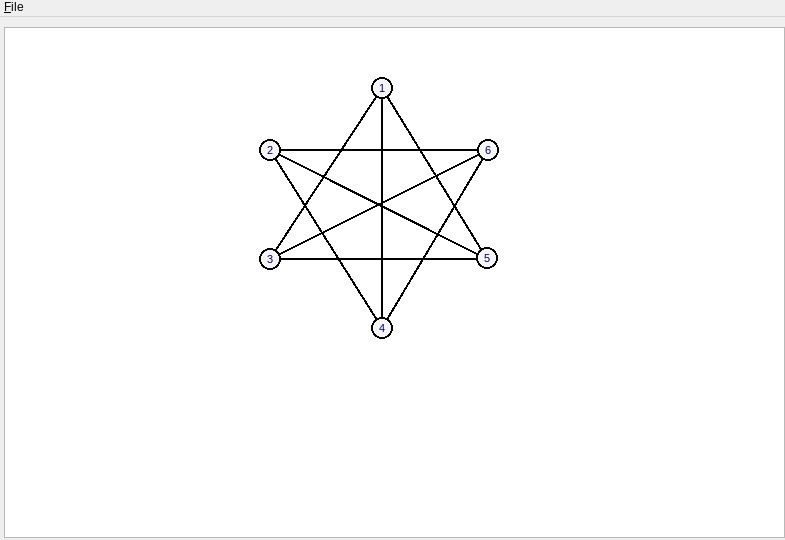}
        \caption{Complement of the six-vertices graph.}
        \label{fig:6v_comp}
    \end{subfigure}
    \caption{A six-vertices graph and its complement.}
    \label{fig:2ab}
\end{figure}

If they are linked by an edge \textit{ab} then, vertices (qubits) \textit{a} and \textit{b} are termed \enquote{adjacent}; in the case of Figure \ref{fig:2v_1e}, vertices 1 and 2 are adjacent.  Furthermore, any and all vertices (qubits) adjacent to vertex \textit{a} form a sub-graph of \textit{G}, termed the \enquote{neighbourhood} of vertex \textit{a} and denoted as $N_a$.  The \textit{complement} of a simple graph, \textit{G} with vertex set \textit{V(G)}, is itself a simple graph, \textit{G'}, of vertex set \textit{V(G)} but with inverted edges, i.e. if vertices \textit{b} and \textit{c} are neighbours of vertex \textit{a} in \textit{G} but \textit{b} and \textit{c} are not adjacent to each other then, they are adjacent in \textit{G'}; see Figures \ref{fig:6v} and \ref{fig:6v_comp}.

It is also possible to restrict complementation to a neighbourhood of \textit{G}, leaving the rest of \textit{G} unaffected.  This \textit{local} complementation is of particular relevance to modelling of quantum algorithms \cite{VDD04,AMD20}.  When used as an operator on \textit{G}, local complementation partly reproduces the effect of measuring a physical qubit, instances of which are demonstrated in relation to Q2Graph in section IV.  Indeed, local complementation may be indispensable to modelling such emerging computational systems as linear-optical \cite{BR05,KRE07,BBGS21} or fusion-based quantum computers \cite{BBB21}, both of which are premised upon MBQC principles.

Specifically in relation to \textit{G}, measuring qubit $\ket\psi_i$ by $\pm\textit{X},\pm\textit{iX},\pm\textit{Y},$ $\pm\textit{iY},\pm\textit{Z},\pm\textit{iZ}$ of $\mathcal{P}$ effectively removes the qubit from further computational input to \textit{G}.  Depending upon the measurement, local complementation of $N_i$ will also occur \cite{HEB04,HDE06}.  Having covered the essentials of \textit{G} in relation to MBQC, it is appropriate to revisit the graph state, $\ket G$ as its own stabiliser state and the restrictions this places on operators of $\mathcal{P}$.

The graph state, $\ket G$, has received considerable attention as a framework for modelling transformations of $\ket\Phi_C$ \cite{VDD04,HEB04,AB06,HDE06,DHW20,BBD09,DNM09}.  A graph state is a quantum state as represented by a (simple) graph.  The reader is reminded that the quantum state $\ket\Phi_C$ is a stabiliser state, restricted to operators of $\mathcal{P}^N$.  The graph state, $\ket G$ is also a stabiliser representing a commutative subgroup, $\mathcal{S}^N$ of $\mathcal{P}^N$ for which,\begin{equation}
    \mathcal{S}^N\coloneqq\{\textit{I},\textit{X},\textit{Y},\textit{Z}\}^N.
\end{equation}
These Pauli operators will leave $\ket G$ at eigenvalue +1 \cite{VDD04,VDD04a}.  Equivalently, $\ket G$ is the simultaneous fixed point of correlation operators, $\textit{K}_j$ entirely determined by \textit{G} \cite{BBD09} such that,\begin{equation}
    \textit{K}_j=\sigma_x^{(j)}  \bigotimes_{\{i,j\} \in \textit{E}} \sigma_z^{(i)},
\end{equation}
and for which $\sigma_x\equiv\textit{X}$ and $\sigma_z\equiv\textit{Z}$.

Two further properties pertaining to these Pauli operators are of note:
\begin{enumerate}
    \item combinations of \textit{X},\textit{Y} and \textit{Z} measurements can replicate the effect of so-called \enquote{Clifford gates}, which include the Hadamard-, phase- and CNOT gates.  By definition, $\ket G$ and hence, $\ket\Phi_C$ can replicate any quantum circuit composed of Clifford gates \cite{RB02,VDD04,AB06}, which has implications for tracking state in $\ket\Phi_C$ both before and after an arbitrary transformation.  Tracking state transformations in $\ket\Phi_C$ is further explored in section V; and
    \item as a corollary of 1), two graph states, $\ket{G}_i$ and $\ket{G}_j$ are termed \textit{local equivalent} iff $\ket{G}_j$ can be obtained from $\ket{G}_i$ by applying  local unitary, $\textit{U}_{\mathcal{S}}$ (i.e. $\textit{U}_{\mathcal{S}}\ket{G}_i=\ket{G}_j$).  The significance of local equivalence lies in the fact that it guarantees continuity of \textit{G} as an abstraction of $\ket\Phi_C$: one can transform $\ket\Phi_C$ to $\ket{\Phi'}_C$ through operations of $\mathcal{S}$ in the certainty of an otherwise undisturbed state space \cite{Sch04}.  
\end{enumerate}
Through replicating the properties of Clifford gates, \textit{G} is expressive and tractable as syntax for encoding algorithms; through local equivalence, $\ket G$ is suitable for optimising an algorithm.  This optimisation technique is known as an algorithm-specific graph and is further explored in section III.

\section{Q2Graph as a modelling framework}
Some comments on the intrinsic design principles of Q2Graph seem timely.  To begin with the use case for Q2Graph, the application is intended to resolve algorithm design problems using classical computing facilities.  In this context, parallels may be drawn with the original use case of the quantum circuit simulator, Quirk\footnote{url: https://algassert.com/2016/05/22/quirk.html, accessed August 2022.} in that there is a limited availability of software to support graph state modelling.  Two further design principles inform the Q2Graph use case, differentiating the application from circuit-based quantum simulators:
\begin{enumerate}
    \item a graph as the algorithm passed to the \enquote{back end}, whether that be another simulator or quantum computer, and
    \item a modelling tool suited to current NISQ computing facilities.
\end{enumerate}
The components and configuration of \textit{G} as well as a restricted set of operators to transform it are central to realising these design principles.

As stated above, a vertex of Q2Graph may signify either a physical or a logical qubit.  The user of Q2Graph must be free to reason about structure or characteristics of \textit{G} as an algorithm without having to account for (quantum) errors and their impact upon state.  Furthermore, only the $\mathcal{S}$ operators, \textit{I},\textit{X},\textit{Y} and \textit{Z} are at the user's disposal (\textit{caveat}, see section IV for a discussion of local complementation as a Q2Graph operation) therefore enforcing the rule that every iteration of \textit{G} is a stabiliser state.  Unequivocally, hardware-side problems of implementing the algorithms it produces are out of scope for Q2Graph.

\begin{figure}[h!]
    \centering
    \includegraphics[width=88mm]{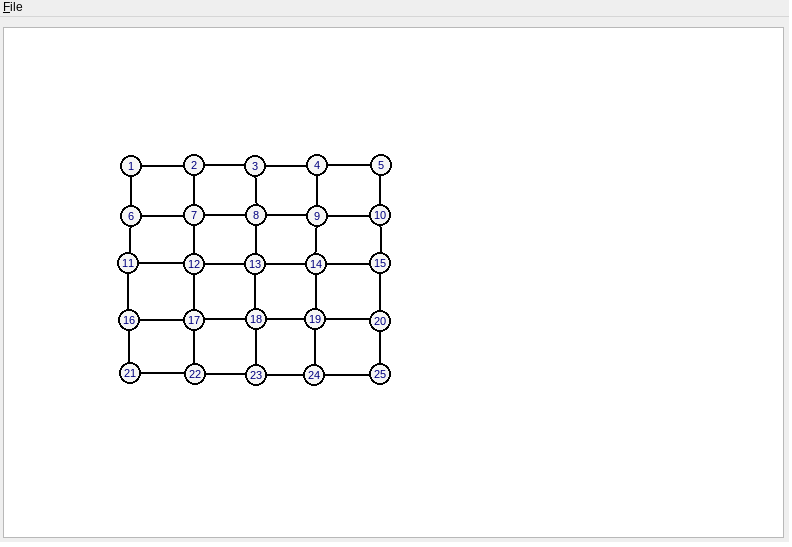}
    \caption{a two-dimensional, 5 * 5 lattice.}
    \label{fig:55_lattice}
\end{figure}

Priority is given to a ready interpretation of \textit{G} and efficiency of reducing it on a classical computer thus, Q2Graph is strictly two-dimensional in its representations of $\ket\Phi_C$ as in the example of Figure \ref{fig:55_lattice}, below.
Indeed, spatial orientation of qubits is largely irrelevant to Q2Graph even though $\ket\Phi_C$ sometimes is a three-dimensional structure \cite{RHG06}.  The only spatial consideration relevant to Q2Graph is minimising the distance of a graph, as a concession to optimisation.  Limiting \textit{G} to a two-dimensional representation is also significant to the concept of the algorithm-specific graph.

Processing a computation through $\ket\Phi_C$ is potentially wasteful of physical qubits because a proportion of the lattice qubits may be superfluous \cite{RB01,RB02}.  In any such instance, care should be taken to minimise the resource requirements of an algorithm.  In this context, it helps to think of a (quantum) computational problem as potentially composite: some elements of the problem may be efficiently soluble with a classical computer leaving only a \enquote{remainder} problem that requires processing with a quantum computing facility.  The algorithm-specific graph (henceforth, ASG) is a design approach to involve classical computing hardware to minimise the qubit count of an algorithm before it is passed to a quantum computing facility.  The idealised ASG breaks down thus:
\begin{enumerate}
    \item In a \textit{classical} computing facility,
    \begin{enumerate}
    \item[(a)] all qubits of $\ket\Phi_C$, except input qubits, are at state $\ket +$,
    \item[(b)] in order, apply $\mathcal{S}$ operators to (a) - recall, these operators can combine to replicate Clifford gates - leaving as a remainder those qubits that require non-Clifford operations (e.g. multi-qubit gates).
    \end{enumerate}
    \item As required, in a \textit{quantum} computing facility,
    \begin{enumerate}
        \item[(c)] execute non-Clifford operations or measurements on any remainder of 1(b).
    \end{enumerate}
\end{enumerate}
 It is important to recognise that \textit{G}, as designed in Q2Graph, is the intermediate representation (Cf. \cite{CBS17,HSS18}) of steps (a) through (c).  Steps (a) and (b) can be thought of as trimming an algorithm of computation that can be efficiently handled on a classical computer.  As stipulated above, Q2Graph only admits transformation of \textit{G} through Pauli measurements hence every step of the computation is a stabilised state (i.e. $\ket G$) and therefore errors are not unwittingly passed from classical computing to quantum computing facilities.  As an approach, the ASG is both an accommodation with the resource limitations of current NISQ technology\footnote{For example, as at December 2021 an upper limit of 127 physical qubits was accessible through IBMQ\_washington (url: https://quantum-computing.ibm.com/services?services=systems\&system =ibm\_washington).} and recognition that facilities of quantum computing and classical computing will coexist and interact, rather than the former supplanting the latter.

\section{Q2Graph: graph operations}
\begin{figure}[h]
    \centering
    \includegraphics[width=88mm]{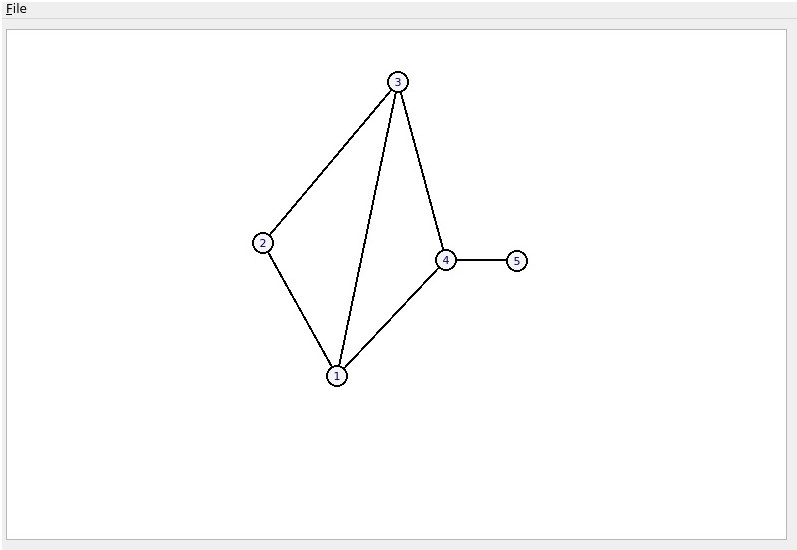}
    \caption{A five-vertices graph.}
    \label{fig:5v}
\end{figure}

Upon opening Q2Graph, the user will see a bounded work canvas with a minimal toolbar.  Most Q2Graph functions are activated via keystrokes while a graph's features are reorganised either through further functions or by means of drag-and-drop mouse functions.  The initial functions are to add or remove a vertex or an edge therefore counting as the backbone to constructing and editing of \textit{G}; an example of a five-vertices graph created by Q2Graph appears as Figure \ref{fig:5v}.
\begin{enumerate}
    \item Add/remove vertex: the user can,
    \begin{itemize}
        \item \textbf{add} a vertex by punching the \enquote{v} key then, left-clicking the mouse at the desired position on the canvas.  At any point, the user may click-and-drag a vertex instance to change its position on the canvas.
        \item \textbf{remove} a vertex by right-clicking the target vertex then, selecting option \enquote{Delete} (or, keystroke \enquote{d}), as per Figure \ref{fig:5v_del}.
    \end{itemize}
    \begin{figure}[h]
        \centering
        \includegraphics[width=88mm]{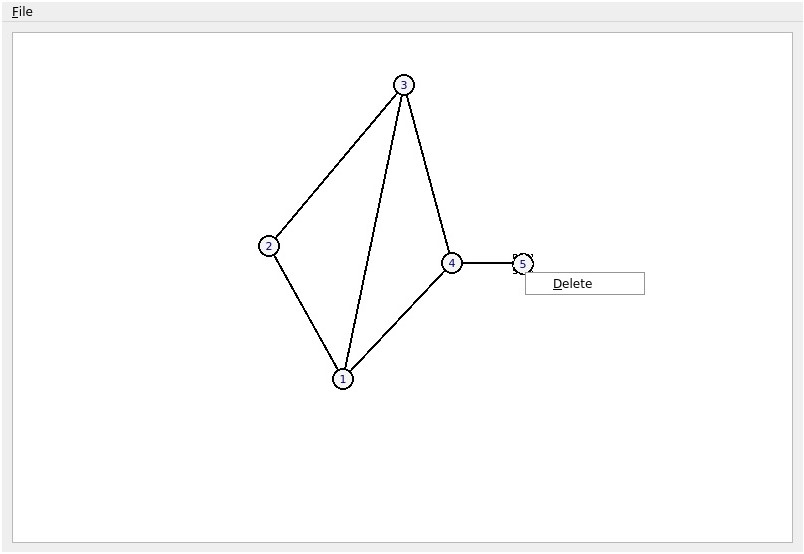}
        \caption{Deleting a vertex of \textit{G}.}
        \label{fig:5v_del}
    \end{figure}
    \item Add/remove edge: an edge, \textit{ab}, signifies an interaction between vertex \textit{a} and vertex \textit{b}.  The user can,
    \begin{itemize}
        \item \textbf{add} an edge by punching the \enquote{e} key then, left-clicking first on (nominated) vertex \textit{a} then on vertex \textit{b}.
        \item \textbf{remove} an edge by right-clicking on the edge itself then, selecting \enquote{Delete} (or, keystroke \enquote{d}), as per Figure \ref{fig:5v_edge_del}.
    \end{itemize}
    As noted above, both a loop edge - an edge that connects a vertex to itself - and multiple edges between two vertices are prohibited as options in Q2Graph.
    \begin{figure}[h]
        \centering
        \includegraphics[width=88mm]{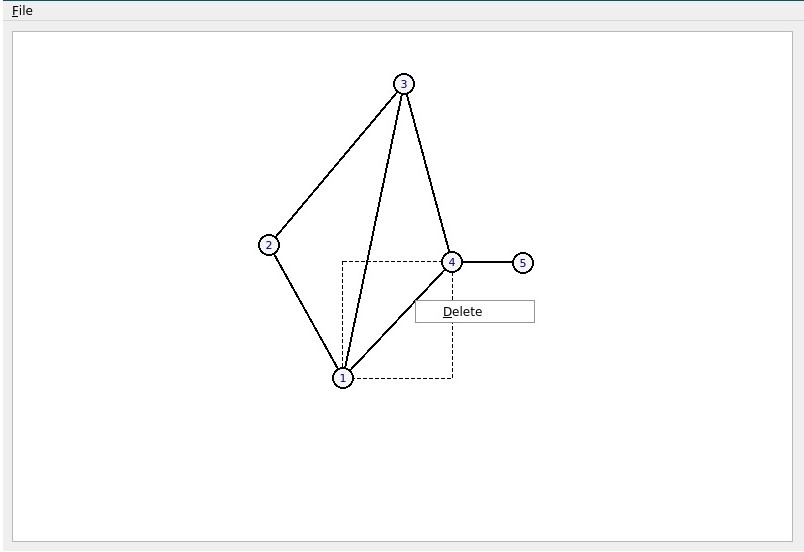}
        \caption{Deleting an edge of \textit{G}.}
        \label{fig:5v_edge_del}
    \end{figure}
    \item Local complementation.  To perform local complementation as a standalone function, punch the \enquote{o} key then, left-click on (nominated) vertex \textit{a}.  Figure \ref{fig:5v_o_lc} is a demonstration of local complementation applied to vertex 4 as it appears in the graph of Figure \ref{fig:5v}.  Strictly speaking this function should be reserved for testing purposes only; if $\ket G$ is to be preserved as part of an ASG then, all transformation should occur via the Pauli operators \textit{X}, \textit{Y} or \textit{Z} (see below).
    \begin{figure}[h]
        \centering
        \includegraphics[width=88mm]{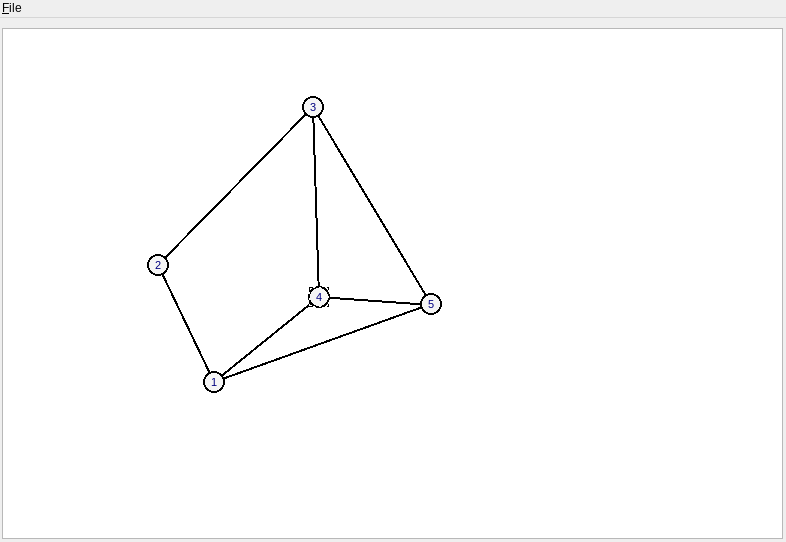}
        \caption{Local complementation applied to vertex 4 of \textit{G}.}
        \label{fig:5v_o_lc}
    \end{figure}
\end{enumerate}

The Pauli measurements (henceforth, LPMs) in Q2Graph - $\sigma_z$ (Z), $\sigma_y$ (Y) and $\sigma_x$ (X) - are operations for transforming \textit{G}.  There is no Undo function to rollback an LPM operation in version 0.1 of Q2Graph although the user may add vertices and edges to restore a configuration or alternatively, \underline{F}ile $>$ \underline{S}ave in JSON format, at any point.  After \cite{HDE06}, each of $\sigma_z$, $\sigma_y$ and $\sigma_x$ respectively transforms \textit{G} thus:
\begin{enumerate}
    \setcounter{enumi}{3}
    \item $\sigma_z$ (Z): to apply this operation, punch the \enquote{z} key then, left-click on (nominated) vertex \textit{a} to remove it and each of its edges from \textit{G}.  Figure \ref{fig:5v_z} is a demonstration of this LPM applied to vertex 1 as it appears in the graph of Figure \ref{fig:5v}; note too how the remaining vertices of Figure \ref{fig:5v_z} have renumbered as a result of the Z-operation, this occurs in Q2Graph as a byproduct of any delete operation.
    \begin{figure}[h]
        \centering
        \includegraphics[width=88mm]{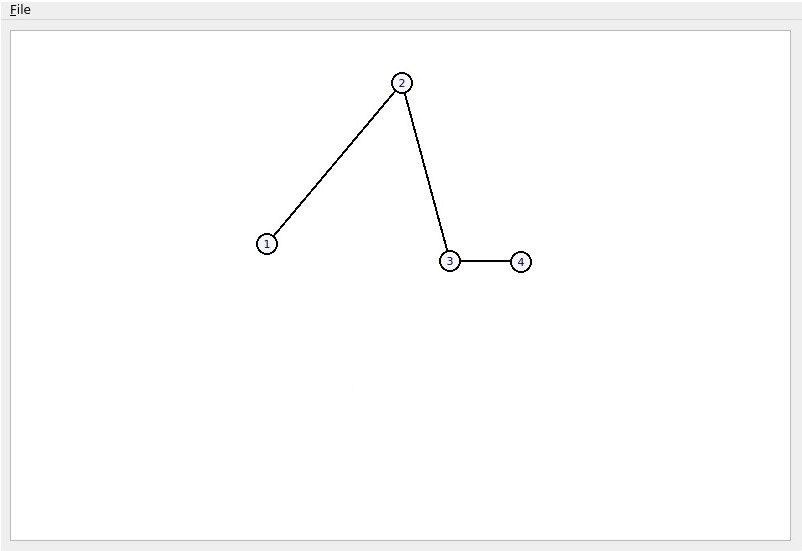}
        \caption{Z-operation applied to five-vertices graph of Figure \ref{fig:5v}.}
        \label{fig:5v_z}
    \end{figure}
    \item $\sigma_y$ (Y): to apply this operation, punch the \enquote{y} key then, left-click on (nominated) vertex \textit{a} to execute local complementation on it; a Z-operation on vertex \textit{a} then completes the Y-operation.  Figure \ref{fig:5v_y} is a demonstration of this LPM applied to vertex 1 as it appears in the graph of Figure \ref{fig:5v}.
    \begin{figure}[h]
        \centering
        \includegraphics[width=88mm]{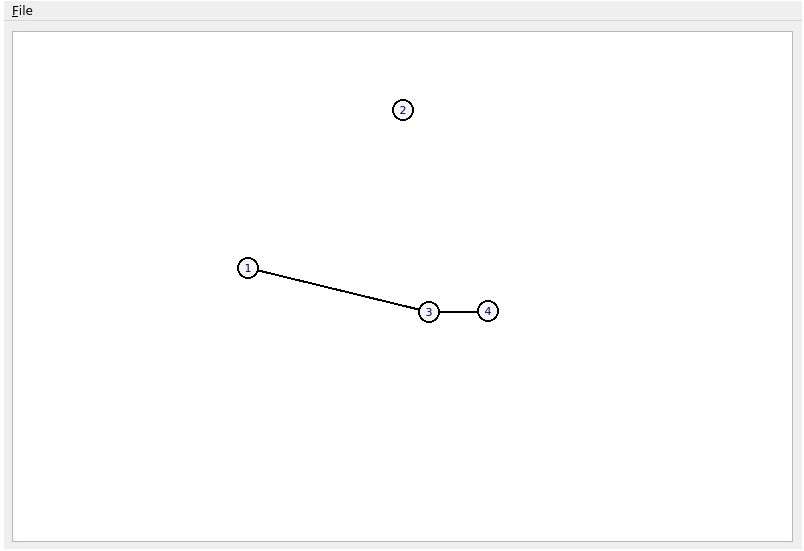}
        \caption{Y-operation applied to five-vertices graph of Figure \ref{fig:5v}.}
        \label{fig:5v_y}
    \end{figure}
    \item $\sigma_x$ (X): is an operation requiring two consecutive left-clicks.  To effect the X-operation, first punch the \enquote{x} key then, left-click on (nominated) vertex \textit{a} to execute local complementation on it.  Figure \ref{fig:5v_x_lc1} is a demonstration of the local complementation of vertex \textit{a}, in this case vertex 1 of the graph in Figure \ref{fig:5v}.  Note how the neighbourhood of vertex 1 is highlighted as a prompt to the user to select the \enquote{special neighbour} vertex \cite{HEB04,HDE06}.  To conclude the operation, select and left-click on the special neighbour vertex, \textit{b} to execute local complementation on it; $\sigma_z$ transformation of vertex \textit{a} then completes the X measurement.  Figure \ref{fig:5v_x_lc2} is a demonstration of these concluding operations, in this instance vertex 4 as it appears in Figure \ref{fig:5v_x_lc1} is the nominated vertex \textit{b}.
    \begin{figure}[h]
    \centering
    \begin{subfigure}{0.4\textwidth}
        \includegraphics[width=\textwidth]{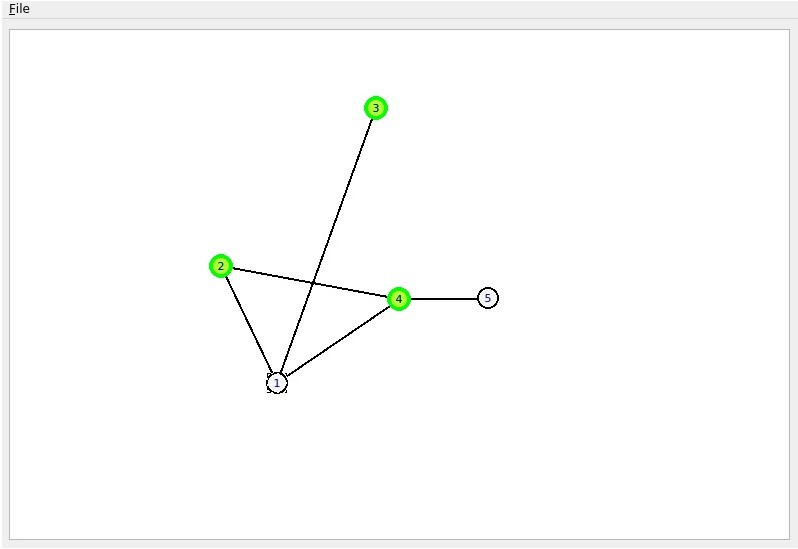}
        \caption{First local complementation as part of the X-operation.}
        \label{fig:5v_x_lc1}
    \end{subfigure}
    \hfill
    \begin{subfigure}{0.4\textwidth}
        \includegraphics[width=\textwidth]{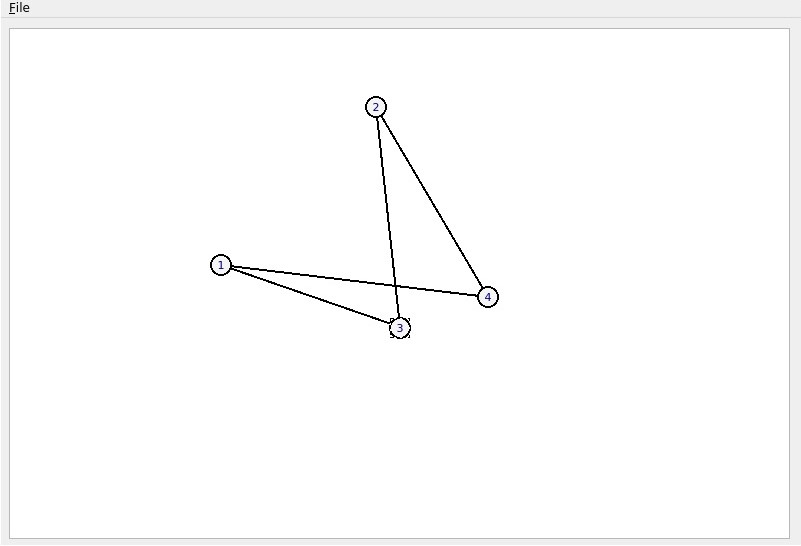}
        \caption{Second local complementation on vertex 4 as part of the X-operation.}
        \label{fig:5v_x_lc2}
    \end{subfigure}
    \caption{X-operation applied to five-vertices graph of Figure \ref{fig:5v}.}
    \label{fig:9ab}
    \end{figure}
\end{enumerate}

\section{Conclusion and next steps}
Q2Graph is an application to resolve quantum problems using classical computing facilities or, more generally, a stable framework for testing the algorithms necessary to an efficient processing of $\ket\Phi_C$.  The user of Q2Graph is able to create and transform a simple graph, \textit{G} by:
\begin{enumerate}
    \item adding or removing a vertex or an edge,
    \item executing local complementation on a target subgraph,
    \item reproducing LPMs - $\sigma_x$, $\sigma_y$, $\sigma_z$ - through functions that combine elements of 1. with 2.
\end{enumerate}
In turn, the restrictions of format and transformation of \textit{G} are the guarantee of $\ket G$ in any algorithm passed to a downstream quantum computing facility.  

What then is planned for Q2Graph version 0.2?  The immediate priority is reliable testing of \textit{G} as an algorithm although there are no obvious means to \enquote{pipe} a graph structure through any simulator that is currently (or freely) available.  As mentioned, the majority of simulators presently available accept only the circuit model as their algorithm.  Furthermore, many simulators are either an SDK, most often tightly-coupled front- and back-end source code, or a PaaS, often a limited licence, proprietary GUI + API combination (Cf. \cite{LFS20}).  Put plainly, it would be a non-trivial undertaking to reconfigure an open-source simulator in order for it to accept a graph as algorithm.  One short-term solution would be to specify a transpiler of graph-to-QASM or graph-to-JSON in order to use a language like Cirq\footnote{url: https://quantumai.google/cirq, accessed August 2022.} or a platform like Qiskit\footnote{url: https://qiskit.org/, accessed August 2022.} as vessels for submitting \enquote{graphs} to a simulator.  While the transpiler approach does present a solution of sorts, it is also potentially tortuous and carries the risk of information loss.  The long-term goal must be extending the Q2Graph toolchain to include a bespoke simulator.

The other priority for Q2Graph version 0.2, closely related to realising a bespoke simulator, is designing a variation on a \enquote{Pauli tracking} facility, as a record of state changes to $\ket\Phi_C$.  Recall, the outcome of measuring qubit $\ket\Phi_C^{\textit{i}}$ is probabilistic so there is a requirement to apply local unitary operators after all measurements are completed, to make $\ket\Phi_C$ a deterministic framework.  Pauli tracking is a log of the requisite corrections induced by probabilistic measurements and thus, acts as a script for the local unitary operators necessary to rectify the affected qubit(s) \cite{Kni05,PDN14,Ter15}.  It is already evident in the case of Q2Graph that a Pauli measurement will execute in full every time and without propagating errors to $\ket\Phi_C$.  It is therefore proposed to use the event of a Clifford operation \enquote{firing} and the \textit{subsequent encoding of state} to $\ket{\hat{\Phi}}_\textit{N}$ (as consistent with (5), above) as a record of an algorithm's outcomes.  A Q2Graph log file would account for all state changes from initialisation through any transformations to the remainder algorithm identified in relation to an ASG.  A log of state change offers obvious value to proofs or in efforts at debugging an algorithm.

\begin{acknowledgments}
The views, opinions and/or findings expressed are those of the author(s) and should not be interpreted as representing the offcial views or policies of the Department of Defense or the U.S. Government.  This research was developed with funding from the Defense Advanced Research Projects Agency [under the Quantum Benchmarking (QB) program under award no. HR00112230007 and HR001121S0026 contracts].
\end{acknowledgments}

\bibliography{q2graph}

\end{document}